\documentclass[aps,prb,twocolumn,superscriptaddress,showpacs,twoside,10pt]{revtex4-1}
\usepackage[utf8]{inputenc}
\usepackage[T1]{fontenc}
\usepackage{amssymb,amsmath}
\usepackage{graphicx}
\usepackage{color}
\usepackage[colorlinks,bookmarks=false,citecolor=blue,linkcolor=blue,urlcolor=blue,backref=no]{hyperref}
\usepackage{bbm}
\usepackage{subfigure}
\usepackage{breakurl}
\let\oldhat\hat
\renewcommand{\vec}[1]{\mathbf{#1}}
\renewcommand{\hat}[1]{\oldhat{\mathbf{#1}}}

\newcommand{\bi}{\begin{itemize}}
\newcommand{\ei}{\end{itemize}}

\newcommand{\be}{\begin{eqnarray}}
\newcommand{\ee}{\end{eqnarray}}

\DeclareMathSizes{10}{9}{7}{6}

\begin{document}
\title{Anisotropic frustrated Heisenberg model on the honeycomb lattice}

\author{Ansgar Kalz}
\email{kalz@theorie.physik.uni-goettingen.de}
\affiliation{Institut für Theoretische Physik, Georg-August-Universität Göttingen, 37077 Göttingen, Germany}
\author{Marcelo Arlego}
\affiliation{Instituto de Física La Plata and Departamento de Física, Universidad Nacional de La Plata, C.C. 67, 1900 La Plata, Argentina}
\author{Daniel Cabra}
\affiliation{Instituto de Física La Plata and Departamento de Física, Universidad Nacional de La Plata, C.C. 67, 1900 La Plata, Argentina}
\author{Andreas Honecker}
\affiliation{Institut für Theoretische Physik, Georg-August-Universität Göttingen, 37077 Göttingen, Germany}
\affiliation{Fakultät für Mathematik und Informatik, Georg-August-Universität Göttingen, 37073 Göttingen, Germany}
\author{Gerardo Rossini}
\affiliation{Instituto de Física La Plata and Departamento de Física, Universidad Nacional de La Plata, C.C. 67, 1900 La Plata, Argentina}

\date{January 6, 2012; revised March 30, 2012}

\begin{abstract}
We investigate the ground-state phase diagram of an anisotropic Heisenberg model on the honeycomb lattice with competing interactions. We use quantum Monte Carlo simulations, as well as linear spin-wave and Ising series expansions, to determine the phase boundaries of the  ordered  magnetic phases. We find a region without any classical order in the vicinity of a highly frustrated point. Higher-order correlation functions in this region give no signal for long-range valence-bond order. The low-energy spectrum is derived via exact diagonalization to check for topological order on small-size periodic lattices.
\end{abstract}


\maketitle

\section{Introduction \label{s:intro}}
The honeycomb lattice has received much attention in recent years because of its relevance to graphene, 
whose electronic structure gives rise to many unusual features.\cite{p:neto09}
However, this two-dimensional bipartite lattice with its two-site unit cell was investigated long before it was realized in a real material.
It is particularly interesting for quantum-mechanical models of strongly correlated electrons since its coordination number  $n=3$ is the lowest allowed in a two-dimensional system.\cite{P:fouet01, B:richter04b}
Hence the influence of quantum fluctuations on the ground-state properties is expected to be more important than, e.g., in the also bipartite square lattice.
Recently, it was found that the half-filled Hubbard model on the honeycomb lattice exhibits a spin liquid state at the border of the metal-insulator transition for an intermediate value of the on-site repulsion $U$.\cite{P:meng10, P:clark11}
Since then the investigation of spin models that can be derived perturbatively from the Hubbard model,\cite{P:yang10, *P:yang11} or stated directly inside the insulating phase, has yielded many interesting features including disordered and valence bond solid phases in the ground-state phase diagram.\cite{P:mulder10, P:cabra10, P:mecklenburg11, P:albuquerque11, P:oitmaa11, P:mezzacapo11} Furthermore, in the context of iridium compounds, systems have been investigated that may be described by frustrated spin models on the honeycomb lattice and exhibit antiferromagnetic order.\cite{P:kimchi11}

Quantum Monte Carlo (QMC) simulations are rather difficult to apply for these spin models due to the sign problem that appears in relation to frustrating quantum spin fluctuations  and thermalization problems accompanying the competition between different ground states.
The latter can be overcome by additional exchange Monte Carlo updates, as was shown in previous work for frustrated spin models\cite{P:melko07} and in particular for an anisotropic $J_1$-$J_2$ Heisenberg model on the square lattice.\cite{P:kalz11,*P:kalz11e,*P:kalz11c}
To avoid the sign problem completely and allow for the investigation of ground-state properties of reasonably large systems it is necessary to lift the frustration for some interactions in such a way that the isotropy of the model is lost.

In the present work we investigate a spin model on the honeycomb lattice, including nearest-, next-nearest-, and third-nearest-neighbor anisotropic Heisenberg interactions, a geometry analyzed in previous works for the isotropic Heisenberg model.\cite{P:fouet01, P:cabra10, P:albuquerque11}
We start from the limit of small quantum fluctuations, where classical antiferromagnetic $S^z$ interactions along a preferred direction are all antiferromagnetic, leading to frustration. However, we choose the quantum fluctuations in the transversal plane to be ferromagnetic and hence nonfrustrating. Such an anisotropic version of the Heisenberg model can be interpreted as a system of hard-core bosons\cite{P:matsubara56} with nonfrustrating kinetic energy and repulsive interactions. In consequence, an application might be realized in optical lattices.\cite{P:lewenstein07, P:struck11}

Applying different techniques, we find the ground-state phase diagram of the anisotropic spin model. We analyze the boundaries of the ordered phases and explore the region between them. Using QMC simulations and exact diagonalization (ED), we identify a finite parameter region where a disordered ground state cannot be excluded. Very similar behavior was observed for the anisotropic $J_1$-$J_2$ Heisenberg model on the square lattice.\cite{P:kalz11} 
The paper is structured as follows. In Sec.~\ref{s:model} we introduce the spin model and its equivalent hard-core boson realization. We briefly present in Sec.~\ref{s:meth} the different methods that will be applied to derive the phase diagram. The results for different limiting cases and from the various approaches are compared and discussed in the main part of the paper in Sec.~\ref{s:results}. The possibility of a spin liquid phase, as well as advantages and drawbacks of the different methods, is discussed in the concluding Sec.~\ref{s:dis}.

\section{Model \label{s:model}}
We consider a spin model on a periodic honeycomb lattice with $N = 2\times(L \times L)$ sites.
The homogeneous anisotropic Heisenberg interactions between spin operators $\vec{S}=(S^x,S^y,S^z)$ at sites $i,j$, separated by a distance called $r$, are given by
\begin{align}
J_{r}\left(\vec{S}_{i},\vec{S}_{j}\right)_{r} & \equiv & J_{r}^{z}S_{i}^{z}S_{j}^{z}+J_{r}^{x,y}\left(S_{i}^{x}S_{j}^{x}+S_{i}^{y}S_{j}^{y}\right)\, ,
\label{eq:interaction}
\end{align}
where $z$ refers to a preferred spin direction, $x$ and $y$ are the transversal directions, and $J_{r}\equiv(J_{r}^{z},J_{r}^{x,y})$ is the anisotropic exchange coupling strength.
The Hamiltonian is given by summing such interactions over all spin pairs on nearest-neighbor ($r=1$), next-nearest-neighbor (NNN) ($r=2$), and third-nearest-neighbor ($r=3$) bonds of the honeycomb lattice:
\begin{align}
H =   \sum_{r=1}^3 \sum_{\langle i,j \rangle_r} J_r \left(\vec{S}_{i},\vec{S}_{j}\right)_r\, .
\label{eq:hamil1}
\end{align}
\begin{figure}[!t]
\begin{center}
\subfigure[\label{f:neelcoll}Ground-state configurations]{\includegraphics[width=0.46\linewidth]{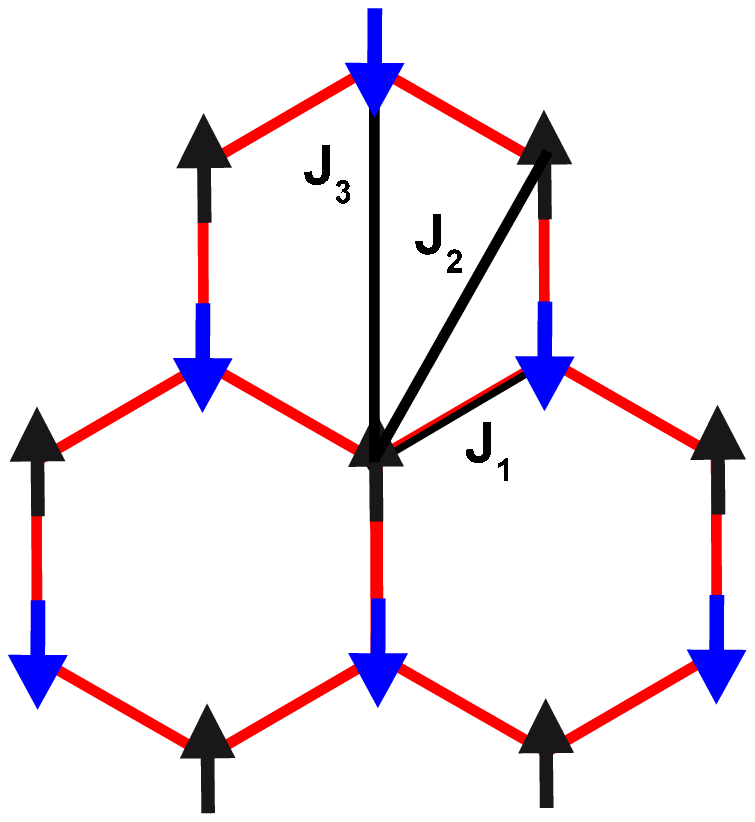}\hspace*{0.04\linewidth}
\includegraphics[width=0.46\linewidth]{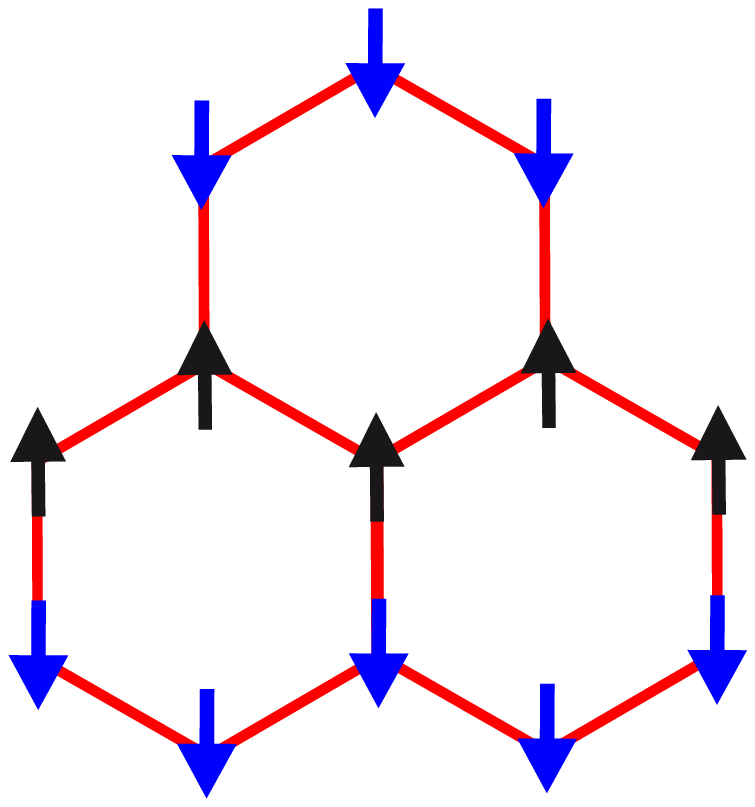}}\\
\subfigure[\label{f:phase_scheme}Schematic phase diagram]{\includegraphics[width=0.96\linewidth]{phase_scheme.eps}}
\caption{\label{f:sketch} (a) Classical ground states: Néel and collinear order on the honeycomb lattice. (b) Schematic phase diagram for the frustrated anisotropic Heisenberg model on the honeycomb lattice. The axes represent the frustration $V = V_2/V_1 = V_3/V_1$ and the relative amplitude of the quantum fluctuations $t = t_r/V_r$. Red circles refer to actual transition points calculated with QMC simulations (see below).}
\end{center}
\end{figure}
Note that $J_1$ and $J_3$ couple sites of the different sublattices (say, $A$ and $B$) of the bipartite honeycomb lattice, while the NNN sites coupled via $J_2$ lie in the same sublattice.
We choose the $S^z$ interactions to be antiferromagnetic, $J^z_r > 0$, such that in the Ising limit ($J^{x,y}_r = 0$) $J_1^z$  and $J_3^z$ favor a N\'eel configuration, with opposite spin orientation in each sublattice [see Fig.~\ref{f:neelcoll}, left].
However, with $J_2^z>0$ a competing interaction is introduced, generating frustration in the model. 
The sign of the exchange terms $J_1^{x,y}$ and $J_3^{x,y}$ does not matter since a sublattice rotation can change both signs simultaneously. However, for the NNN exchange a negative (ferromagnetic) sign has to be chosen to avoid the sign problem in the QMC simulations. For convenience we set all $J^{x,y}_r < 0$ so that the fluctuations are ferromagnetic and hence nonfrustrating.

The present anisotropic Heisenberg model, for spin $S=1/2$, can also be understood as a model of hard-core bosons.
For convenience we first fix an energy scale $J$, defining dimensionless exchange couplings
$J_r^z/ J=V_r $ and $J_r^{x,y}/J = 2\,t_r $.
We rewrite the spin model in terms of ladder operators $S_i^\pm= S_i^x \pm i S_i^y$,
\begin{align}
H /J  =  \sum_{r=1}^3  \sum_{\langle i,j \rangle_r}\left(  V_rS^z_i S^z_j +  t_r (S^+_iS^-_j + S^-_iS^+_j) \right)
\label{eq:hamil2}
\end{align}
and map spin operators onto hard-core bosons by (cf., e.g., Ref.~\onlinecite{P:matsubara56})
\begin{align}
&S_i^+ \rightarrow b_i^{\dagger} \,,\quad S_i^- \rightarrow b_i^{\phantom{\dagger}} \,,\quad S_i^z \rightarrow n_i-1/2\,,  
\label{eq:mapping}
\end{align}
where 
\begin{align}
 &n_i = b_i^{\dagger}b_i^{\phantom{\dagger}} \quad \text{and} \quad (b_i^{(\dagger)})^2 = 0 \,,\label{eq:constraint}
\end{align}
so that the Hamiltonian~\eqref{eq:hamil2} maps to
\begin{align}
H_{\text{boson}}/J &=  \sum_{r=1}^3  \sum_{\langle i,j \rangle_r} \left(t_r(b_i^{\dagger}b_j^{\phantom{\dagger}} + b_j^{\dagger}b_i^{\phantom{\dagger}})+ V_r n_i n_j \right)\label{eq:boson} \nonumber\\
  & -\frac{3}{2}(V_1+2V_2+V_3) \sum_{i}n_i + \frac{3}{8}(V_1+2V_2+V_3)N\, . 
\end{align}
Here the (negative) $t_r$ describe nonfrustrating hopping of the hard-core bosons up to third-nearest neighbors and the (positive) $V_r$ describe repulsion up to the same range. In particular we are interested in the zero magnetization case, mapping to a half-filled lattice, i.e., $\langle n_i \rangle = 1/2$. In this case the last line in Eq.~\eqref{eq:boson} only yields constant terms that will be dropped in all following considerations.

For the remainder of this work, all three interactions will have the same anisotropy ratio $J_r^{x,y}/J_r^z=2\,t<0$, 
the relative strength of the interactions is set to $V_2/V_1=V_3/V_1=V>0$, and the scale is set to $J_1^z=J$ (i.e., $V_1=1$, implying $t_1=t$ and $t_2=t_3=Vt$).
Thus the parameter space is reduced to a two-dimensional area, which we will investigate for positive $V$ and negative $t$.

Ground-state configurations are readily obtained in some limiting cases and a schematic phase diagram is 
anticipated in Fig.~\ref{f:phase_scheme}.
In the classical Ising limit ($t \to 0$) of the spin model in Eq.~\eqref{eq:hamil2}, two different ground states occur:
for $V < 1/2$ a Néel state [Fig.~\ref{f:neelcoll}, left], with energy
\begin{align}
E_{\text{Néel}}/J&= \tfrac{3}{2}(V-1) S^2 N \,, \label{eq:Neel_grounden}
\end{align}
and for $V > 1/2$ a collinear state [Fig.~\ref{f:neelcoll}, right], with energy
\begin{align}
E_{\text{coll}/J}&= \tfrac{1}{2}(1-5V) S^2 N\,.  \label{eq:grounden}
\end{align}

At the critical point $V=1/2$ the two ground states compete and the transition temperature is suppressed to zero.
The degenerate ground-state manifold at this point is expected to give rise to interesting phenomena for nonzero quantum fluctuations ($t<0$) in the direct vicinity of the critical point.

The limit of large values of $|t|$ can be easily understood too as the noncompeting ferromagnetic interactions yield a ferromagnetic correlation in the $x$-$y$ plane [see Fig.~\ref{f:phase_scheme}]: 
The ground state will spontaneously break the rotation symmetry in the $x$-$y$ plane, still with vanishing magnetization in the $z$ axis. The anisotropic Heisenberg interactions \eqref{eq:interaction} can also be understood in terms of Ising interactions between the, say, $S^x$ spin components, regarding the interactions between the other spin components $\tilde{S}^\pm=S^y\pm i S^z$ as (not negligible) spin fluctuations:
\begin{align}
H =& \sum_{r=1}^3 \sum_{\langle i,j \rangle_r} 2t_r S_i^x S_j^x \nonumber \\
&+ \sum_{r=1}^3 \sum_{\langle i,j \rangle_r} \tfrac{2t_r-V_r}{4} (\tilde{S}_i^+\tilde{S}_j^+ + \text{H.c.}) \nonumber \\
&+ \sum_{r=1}^3 \sum_{\langle i,j \rangle_r} \tfrac{2t_r+V_r}{4} (\tilde{S}_i^+\tilde{S}_j^- + \text{H.c.})\,.
\label{eq:hamil_sx}
\end{align}
Then, for large $|t|$ a ferromagnetic product wave function in the $S^x$ basis serves as variational ansatz (VA), with energy
\begin{align}
E_{\text{ferro}}/J = 3\,t (1+3V)S^2 N \,. \label{eq:va_ferro}
\end{align}

In the bosonic language, the limits described above match density waves at small $|t|$ and a superfluid phase at large $|t|$,\cite{P:wessel07} 
corresponding to the condensation of ferromagnetic magnons.\cite{P:bloch31, P:hoehler50, P:matsubara56}
Note that Néel and collinear phases are only exact ground states for $t=0$  or large $S$, i.e., in the classical limit, while the bosonic treatment is valid for $S=1/2$ and will be studied at finite $t$.
One should then expect that fluctuations will reduce the magnetic order.
To determine the stability boundaries of these phases in the $V$-$t$ plane [sketched in Fig.~\ref{f:phase_scheme}] and to analyze the intermediate region of the phase diagram, we apply various methods, which are briefly introduced in the following section.

\section{Methods \label{s:meth}}
In this section we will briefly summarize the methods that we have employed; results will be presented in Sec.~\ref{s:results}. The reader who is not interested in technical details may skip this section.

The series expansion (SE) method will be applied in the limit of small fluctuations to calculate energies and estimate the phase boundaries of the antiferromagnetic phases. The same holds for the derivation of linear spin waves (LSWs), which will also be applied for the variational ferromagnetic state. The QMC simulations are employed for the whole phase diagram to calculate various order parameters and results for the energies will be compared to the other methods. In addition, for a particular parameter set, ED will be performed to calculate the low-energy spectrum.

\subsection{Series Expansion \label{ss:meth_SE}}
We have analyzed perturbatively the Hamiltonian~\eqref{eq:hamil2} starting from the Ising limit ($t_r=0, r=1,\ldots, 3$), around classical Néel and collinear phases. Using standard (Rayleigh-Schrödinger) perturbation theory\cite{P:kogut79} on finite lattices, we have obtained analytic expressions for the ground-state energy up to $\mathcal{O}(t_r^4)$ around the two mentioned classical phases.
We have performed these calculations on finite lattices, large enough to avoid finite-size effects at this order, using computer algebra software for the implementation.\cite{P:reduce} 
The method is straightforward but usually limited to low orders due to the memory constraints imposed by the need of larger lattices to achieve higher orders. Although our expansions are only fourth order, they have the advantage of providing the coefficients of the expansions in a closed analytical form. This allowed us to determine a first-order critical line between Néel and collinear phases, as discussed in the following sections.

Very recently, Oitmaa and Singh\cite{P:oitmaa11} analyzed the isotropic Heisenberg model applying a linked-cluster formalism\cite{B:oitmaa} and calculated numerically series up to eighth order. 
For such a purpose an Ising model is perturbed with an auxiliary anisotropy parameter that in the end has to tend to one to recover the isotropic Hamiltonian. This parameter is directly related to the couplings $t_r$ in Eq.~\eqref{eq:hamil2}; thus their results are useful in analyzing our model. We use here their numerical eighth order series for two purposes: First, we use the ground-state energy results to estimate a range of validity of our series; second and more importantly, we use the series provided for order parameters to determine critical points in the phase diagram of our model.

\subsection{Linear Spin Waves \label{ss:meth_LSW}}
The phase boundaries of the conventional magnetic phases can be estimated for large spin $S$, using a $1/S$ LSW expansion of the model in Eq.~\eqref{eq:hamil2}. 
To this end one selects a classical spin configuration to set local Cartesian frames with $\check{z}_{i}$ along the classical direction 
and transversal components $\check{x}_{i}$ and $\check{y}_{i}$ and represents spins in terms of Holstein-Primakoff bosons $a_{i}$ satisfying $\left[a_{i},a_{j}^{\dagger}\right]=\delta_{ij}$.\cite{P:holstein40} 
The spin component along the local classical direction is represented by
\begin{align}
\tilde{S}_{i}^{z}=\vec{S}_{i}\cdot\check{z}_{i}=S-n_{i},
\end{align}
where $n_{i}=a_{i}^{\dagger}a_{i}$, while transversal fluctuations $\tilde{S}_{i}^{\pm}=\vec{S}_{i}\cdot\check{x}_{i}\pm i\vec{S}_{i}\cdot\check{y}_{i}$ are represented by
\begin{align}
\tilde{S}_{i}^{+}  = & \left(\sqrt{2S-n_{i}}\right)a_{i},\\
\tilde{S}_{i}^{-}  = & a_{i}^{\dagger}\left(\sqrt{2S-n_{i}}\right).
\end{align}
These equations ensure the correct spin component commutation relations as well as $-S\leq\langle\tilde{S}_{i}^{z}\rangle\leq S$. 
The LSW expansion is then obtained by expanding the Hamiltonian up to second order in $a_{i}^{(\dagger)}$; 
while this procedure usually breaks SU(2) invariance, in the present case that symmetry is already explicitly broken by the anisotropy.

The form of the bosonic action depends on the selected classical configuration. In the present case we have explored the N\'eel and collinear configurations as well as a ferromagnetic phase in the $x$-$y$ plane.
In these three configurations the local classical directions at different sites are either parallel or antiparallel, simplifying the computations. 
For classically parallel interacting sites the interaction terms in Eq.~\eqref{eq:hamil2} are expanded as
\begin{align}
V_{r}S^{2}-V_{r}S\left(n_{i}+n_{j}\right)+2St_{r}\left(a_{i}^{\dagger}a_{j}+a_{j}^{\dagger}a_{i}\right),
\end{align}
while for antiparallel local frames they read
\begin{align}
-V_{r}S^{2}+V_{r}S\left(n_{i}+n_{j}\right)+2St_{r}\left(a_{i}^{\dagger}a_{j}^{\dagger}+a_{i}a_{j}\right).
\end{align}
The quadratic Hamiltonian contains anomalous terms in the fluctuations and must be diagonalized by a Bogoliubov transformation.
Within each momentum mode, the Bogoliubov transformation is possible if the Hamiltonian quadratic form has only positive eigenvalues.
We searched such regions in the $V$-$t$ plane and after checking that fluctuations do not destroy the classical order we adopted the possibility of diagonalizing the Hamiltonian as a criterion for stability of the classical phase. Notice that the ferromagnetic phase shows a Goldstone mode at zero momentum, related to the explicit selection of a classical direction in the $x$-$y$ plane; in this phase, stability is achieved if the remaining modes are positive.

Within each phase, we computed the ground-state energy and the average magnetization along the classical directions as order parameters. In the regions where more than one phase is stable, we select the one with lowest ground-state energy.
The results are shown in Sec.~\ref{s:results}.

\subsection{Quantum Monte Carlo \label{ss:meth_QMC}}
The simulation of frustrated spin models using QMC simulations is usually limited by the sign problem that occurs for competing quantum fluctuations. However, for the present model we choose the interactions along the $S^{xy}$ direction of the spin operators to be ferromagnetic and hence have negative amplitudes $t_r$ for the quantum fluctuations. 
This maps onto bosonic hopping integrals [see Eq.~\eqref{eq:boson}], which yield no sign problem in QMC simulations. Thus we can simulate the model for all parameters. Nevertheless, the remaining frustration in the $S^z$ direction of the spin interactions yields a competition between different ground states for $V\approx 1/2$. Exactly at the critical point the classical ground state shows a large degeneracy of linear order $\sim 2^{2\,L}$ and this results in freezing and thermalization problems for simulations of the quantum model at low temperatures.

The QMC simulations were performed at finite temperatures using the implementation of the stochastic series expansion\cite{P:sandvik91} of the ALPS project.\cite{P:alet05, P:ALPS05, P:ALPS07}
The use of directed loops\cite{P:sandvik02} in the update step ensures a reliable scan of the whole phase space even for the present model where several  different operators can act on the same site of the lattice. 
To overcome the freezing problems that occur due to the degenerate ground-state manifold at $V=1/2$ we used an additional exchange Monte-Carlo update\cite{P:hukushima96, P:hansmann97, P:melko07} that allows for a better thermalization of the simulation. Since in particular in the vicinity of the critical point $V\approx 1/2$ low temperatures are necessary to gain insight into the ground-state properties of the system, simulations were run in parallel on large-scale computer clusters.

\subsection{Exact Diagonalization \label{ss:meth_ED}}
To gain further insight into the spectrum of the model we also applied exact diagonalization techniques.
This is done for finite lattices up to $N=34$ sites and only for certain parameters where SE and LSWs are not applicable.
In particular we performed Lanczos iterations to calculate the lowest eigenvalues of the system using an implementation by Jörg Schulenburg.\footnote{The code for the \uppercase{Spinpack} is available at \url{http://www.ovgu.de/jschulen/spin}.}
The finite-size analysis of the behavior of the lowest excited states allows for a characterization of the underlying ground state.

\section{Results \label{s:results}}
In this section we present and compare results obtained by the methods described above  for three different cases of interest.
For small fluctuations the stability of the antiferromagnetic states will be analyzed by means of SE, LSWs and QMC simulations.
In the case of large fluctuations the emergence of long-range ferromagnetic correlations in the $x$-$y$ plane is tested by means of LSWs and QMC simulations through the spin stiffness of the $S^z$ component.
An intermediate region in the vicinity of the critical point $V = 1/2$ will be investigated via QMC calculations of higher-order correlation functions and interpretation of the low-energy spectrum from ED.

\subsection{Ising limit \label{ss:res_ising}}
The Ising limit is given by setting all quantum fluctuations $t_r=0$ and exhibits two antiferromagnetic ground states for $V_2=V_3$ that were described in Sec.~\ref{s:model}. For small values of $|t_r|$ the quantum-mechanical ground states are expected to consist of those classical states plus some quantum fluctuations that reduce the overall energy and order parameters.
\begin{figure}[!t]
\includegraphics[width=0.96\linewidth]{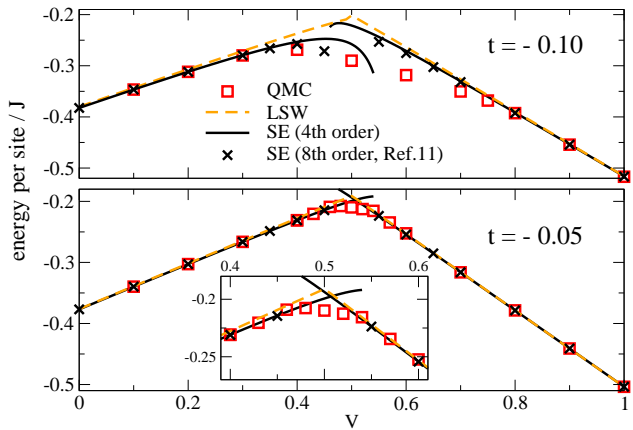}
\caption{\label{f:anti_energies} Comparison of energies calculated from the different methods introduced in Sec.~\ref{s:meth}, at small quantum fluctuations $t=-0.10$ (top) and $t=-0.05$ (bottom). For the SE two different calculations up to fourth order (this work) and up to eighth order (Ref.~\onlinecite{P:oitmaa11}) are shown. In the direct vicinity of the critical point $V=1/2$ both expansions become rather unreliable due to an increasing number of divergences. The LSW results underestimate the influence of quantum fluctuations, as also observed for larger $|t|$.}
\end{figure}

A comparison of the overall energies from SE, LSWs, and QMC simulation is given in Fig.~\ref{f:anti_energies} for fluctuations governed by $t=-0.05$ and $-0.10$. The agreement for small $V<0.45$ and large $V>0.7$ is very good. Only in the intermediate regime can discrepancies be observed, which will be discussed below.

From our fourth-order SE calculations we evaluated ground-state energies for the Néel ($V < 1/2$) and collinear states ($V>1/2$). 
Series for these two states are given in the Supplemental Material.\footnote{See Supplemental Material at the \url{http://arxiv.org/src/1201.1505/anc} for analytic series up to fourth order.} These results agree with the recent work by Oitmaa and Singh\cite{P:oitmaa11} up to the given order.

\begin{figure}[!t]
\includegraphics[width=\linewidth]{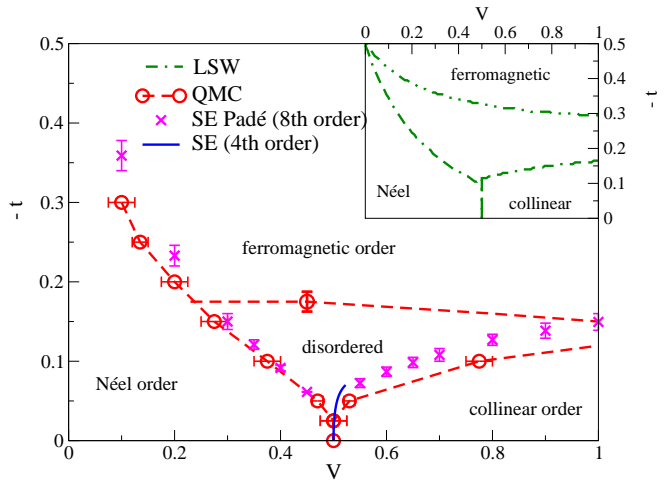}
\caption{\label{f:phase} Ground-state phase diagram for the anisotropic Heisenberg model on the honeycomb lattice by means of QMC simulations (red circles and interpolation by red dashed lines). The Ising ground states survive separated by a direct first-order transition for small fluctuations $|t|$. Only for values of $|t|>0.175(25)$ can ferromagnetic order in the $x$-$y$ plane be detected. 
The solid blue line represents the first-order transition line between the antiferromagnetic states [Eq.~\eqref{eq:SE_4order}], determined from fourth order SE. Magenta crosses represent phase boundaries determined by the condition of vanishing order parameters, provided as eighth order SE (Ref.~\onlinecite{P:oitmaa11}) (see the text for more details).
Inset: The LSWs (green dash-dotted lines) yield phase transition lines due to stability arguments and a comparison of energies between antiferromagnetic states. The overall scenario is very similar. 
}
\end{figure}

The LSW approximations yield comparable results for the energies that are shown in Fig.~\ref{f:anti_energies} for $t=-0.05$ and $-0.10$, computed on a lattice with $2\times 10^{4}$ sites. The intrinsic breakdown of the method for a particular starting configuration as described in Sec.~\ref{ss:meth_LSW} provides an estimate of the upper phase boundary for both antiferromagnetic states: The N\'eel configuration is stable for $-t<\frac{1-V}{1+3V}$ and the collinear one is stable for $-t<-\frac{1-5V}{1+11V}$. 
Both antiferromagnetic configurations support LSW fluctuations for $V \approx 1/2$, where the phase boundary is estimated by energy comparison. The three approximate LSW boundaries are plotted in the inset of Fig.~\ref{f:phase}.

\begin{figure}[!t]
\subfigure[\label{f:neel}Néel order at $V = 0.2$ and $t =-0.1$] {\includegraphics[width=0.96\linewidth]{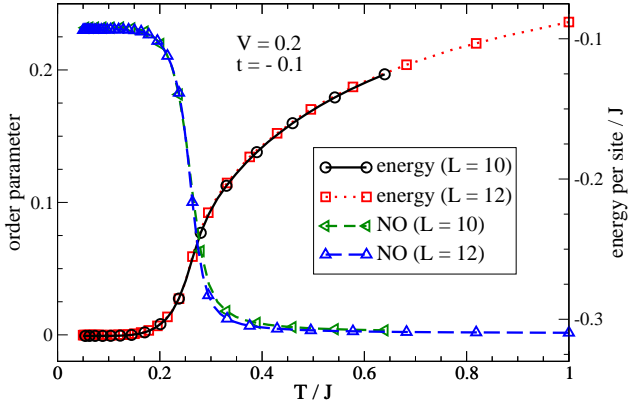}}
\subfigure[\label{f:ferr}ferromagnetic order at $V = 0.45$ and $t =-0.5$] {\includegraphics[width=0.96\linewidth]{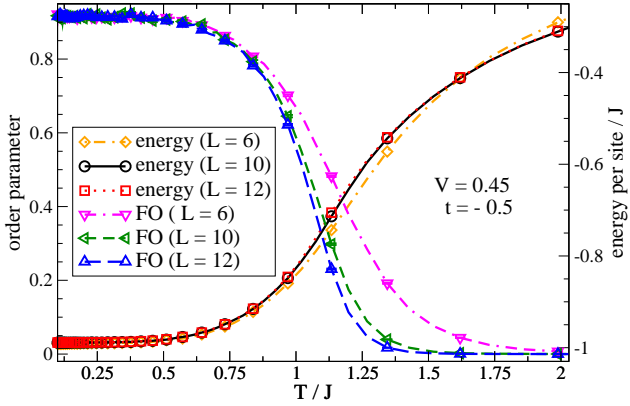}}
\caption{\label{f:orderparameter} Evolution of QMC results for energies and magnetic order parameters shown for decreasing temperatures in two different points of the phase diagram, exhibiting (a) N\'eel order (NO) and (b) ferromagnetic order (FO), respectively.}
\end{figure}

Our most accurate results were obtained by extensive QMC simulations for the energies and magnetic order parameters.
To identify the regions with different magnetic orders we calculate the structure factors for the Néel and collinear order.
The Cartesian wave vector is given by $\vec k = (0,0)$ and antiparallel spins on the $A$ and $B$ sites of the unit cell for the Néel state, i.e., each sublattice is ferromagnetically ordered but they are aligned antiparallel to each other. The collinear state is sixfold degenerate with three wave vectors: $\vec k = \tfrac{\pi}{\sqrt{3}}(\sqrt{3},1)$ and $\vec k = \tfrac{\pi}{\sqrt{3}}(\sqrt{3},-1)$ with $A$ and $B$ parallel and $\vec k = \tfrac{2\pi}{\sqrt{3}}(0,1)$ with $A$ and $B$ antiparallel. Additionally all spins can be flipped in the ordered states, which gives a twofold degeneracy. An example of the behavior of the order parameter as a function of temperature is shown in Fig.~\ref{f:neel} for the Néel state.

As expected from SE for small $|t|$, we obtain a direct transition between the two  antiferromagnetic states that is probably of first order since both states exhibit different symmetries. From QMC simulations we observe that this transition line splits into two for a small value of $0.025 < |t| < 0.05$ and a new ground state emerges in between. 

\subsection{Ferromagnetic limit \label{ss:res_ferro}}
In the opposite limiting case with large  $|t|$ the situation is not as simple as for the Ising limit. Linear spin waves can be expanded around the product state that yields the energy given in Eq.~\eqref{eq:va_ferro} and again the breakdown will yield an estimation of a lower phase boundary. However, the coupling strength of the quantum fluctuations, scaling with $|t|$,  is not small compared to the Ising exchange and the position of this phase boundary is not very reliable. In the inset of Fig.~\ref{f:phase} we show the ferromagnetic phase boundary obtained by LSWs  in a large lattice, well fitted by $-t>\frac{1+1.83V}{1+4V}$. Then LSWs cannot be applied in the unidentified region in the inset of Fig.~\ref{f:phase}. The energies obtained by this calculation are compared below to the results of QMC simulations.

\begin{figure}[!t]
\subfigure[\label{f:V_scan} fixed $t=-0.05$ and varying frustration $V$, $L=12$]{\includegraphics[width=\linewidth]{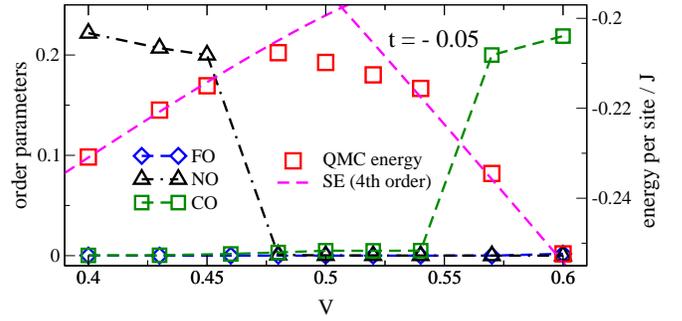}}
\subfigure[\label{f:t_scan} fixed $V=0.45$ and varying fluctuations $t$, $L=12$]{\includegraphics[width=\linewidth]{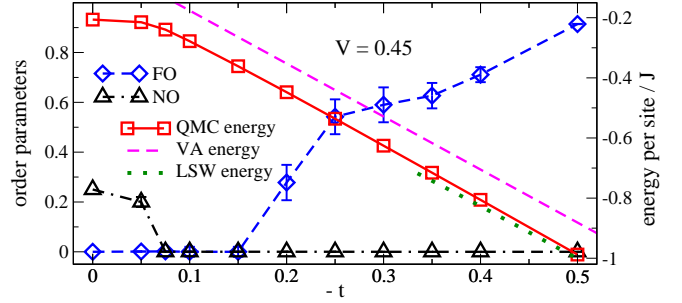}}
\caption{\label{f:scan} Energy and order parameters (FO, ferromagnetic in-plane order; NO, Néel order; CO, collinear order) shown for (a) a horizontal cut through the phase diagram (Fig.~\ref{f:phase}) at $t=-0.05$ (classical order is absent for $0.48 \lesssim V \lesssim 0.54$) and (b) a vertical cut at $V=0.45$ (no finite order parameter for $0.075 \lesssim -t \lesssim 0.15$).}
\end{figure}

The appropriate order parameter in the QMC simulations for the ferromagnetic state is given by the spin stiffness, which can be estimated on behalf of the winding number within the QMC algorithm.\cite{P:matsubara56,P:pollock87} As an example, the convergence of energies and this order parameter is shown in Fig.~\ref{f:ferr}. Careful calculations of this observable for the not antiferromagnetically ordered regions of the phase diagram show only a nonvanishing signal for $|t| > 0.15$. Figure~\ref{f:scan} shows two different parameter scans. In Fig.~\ref{f:V_scan} order parameters and energies from QMC simulations and SE are shown for $t=-0.05$ and varying frustration $V$ from N\'eel to collinear behavior: A finite region without any magnetic order is identified, which also explains the discrepancy of the energies shown here and in Fig.~\ref{f:anti_energies}. In Fig.~\ref{f:t_scan} a similar scan is presented, here for fixed $V=0.45$ and $t$ varying from N\'eel to ferromagnetic behavior, where the QMC energy is also compared with the classical variational ansatz [Eq.~\eqref{eq:va_ferro}] and LSW calculations. The agreement with the LSW energy is remarkably good.

As a result from the QMC, SE and LSW calculations we show a phase diagram for the $V$-$t$ parameter space in Fig.~\ref{f:phase}. The three methods uncover a finite region without magnetic order around the critical point $V=1/2$ and finite fluctuations $-t$, which will be analyzed in the following section. 
The Ising ground states survive separated by a direct first-order transition for small fluctuations $|t|$. Only for values of $|t|>0.175(25)$ is ferromagnetic order in the $x$-$y$ plane detected by means of QMC simulations.
A direct comparison of energies for $V \approx 1/2$ yields a phase transition line between the two antiferromagnetic states. 
We give here the approximate result as a function $V(t)$, obtained by equating
our fourth order SE of energies for both Néel and collinear states and expanding up to linear order in $V$, around $V=1/2$:
\begin{align}
V(t) = \frac{1278676\,t^4-69750\,t^3+6300\,t^2-225}{2665577\,t^4-148050\,t^3+13950\,t^2-450}\,. \label{eq:SE_4order}
\end{align}
This function is shown as solid blue line in the ground-state phase diagram in Fig.~\ref{f:phase}.

As we have mentioned, Oitmaa and Singh\cite{P:oitmaa11} have presented SE data for the magnetic order parameters, i.e., Néel and collinear magnetization, that can be directly translated to our model. We used these data to determine the upper phase boundaries of the antiferromagnetic phases by detecting the point in $V$-$t$ space where order parameters (more precisely, their Padé approximants) vanish. The result is also shown in the phase diagram (Fig.~\ref{f:phase}, magenta crosses). The corresponding error bars are confidence limits obtained by considering the dispersion of predicted critical points for different Padé approximants. 
Note that the antiferromagnetic phase boundaries estimated by SE are generally slightly above the numerical QMC results. A possible explanation is the presence of first-order transitions at these phase boundaries. 

The LSWs (green dash-dotted lines in the inset of Fig.~\ref{f:phase}) yield phase transition lines as described in Sec.~\ref{ss:meth_LSW}.
From the Monte Carlo simulations, the phase diagram is obtained by identifying the regions where the different order parameters show finite signals (red circles in Fig.~\ref{f:phase}). As shown in Fig.~\ref{f:scan}, this leaves a finite region with a ground state that shows no magnetic order. Since the energies calculated by LSWs were slightly above the values of QMC simulations and SE, for increasing $-t$ it is not surprising that the phase boundaries are shifted to higher values as well. 

\subsection{Intermediate regime \label{ss:res_inter}}
For the intermediate regime we find similar behavior as for the square lattice,\cite{P:kalz11} {\em i.e.}, only small signals appear in the spin stiffness for small lattices and intermediate temperatures that scale to zero for larger lattices.
The estimation of the second critical value for $t(V)$, at which ferromagnetic order arises, is rather difficult due to the two-dimensional parameter space and the finite-size problems in the ferromagnetic order parameter. In Fig.~\ref{f:phase} we show a rough estimate for the phase transition line separating the disordered state and the ferromagnetic phase as a red dashed line.
\begin{figure}[!t]
\subfigure[\label{f:neelcorr}Néel state at $V=0.2$ and $t=-0.1$ ($T=0.05~J$)]{\includegraphics[width=0.96\linewidth,height=0.72\linewidth]{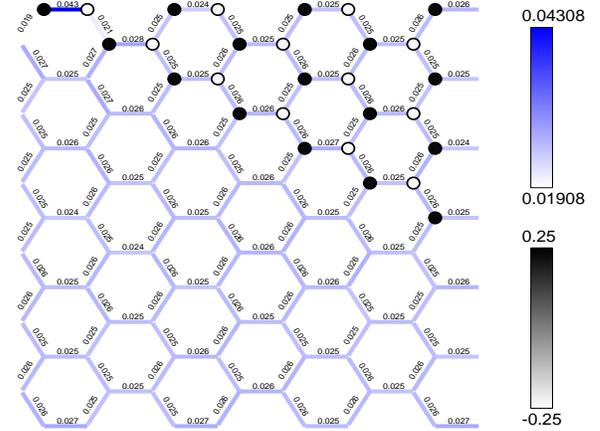}}
\subfigure[\label{f:supercorr}ferromagnetic state at $V=0.45$ and $t=-0.5$ ($T=0.1~J$)]{\includegraphics[width=0.996\linewidth,height=0.72\linewidth]{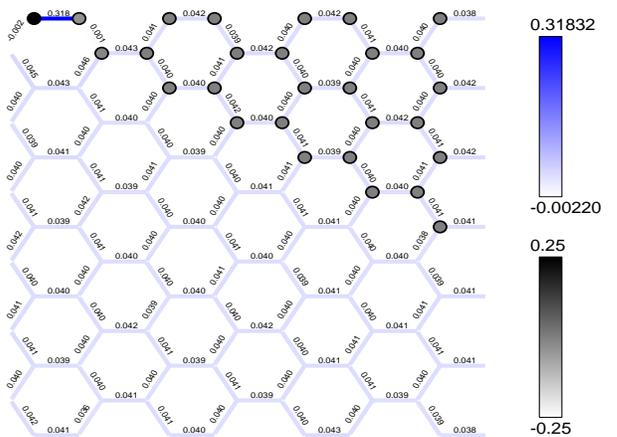}}
\subfigure[\label{f:discorr}disordered state at $V=0.45$ and $t=-0.1$ ($T=0.05~J$)]{\includegraphics[width=0.96\linewidth,height=0.72\linewidth]{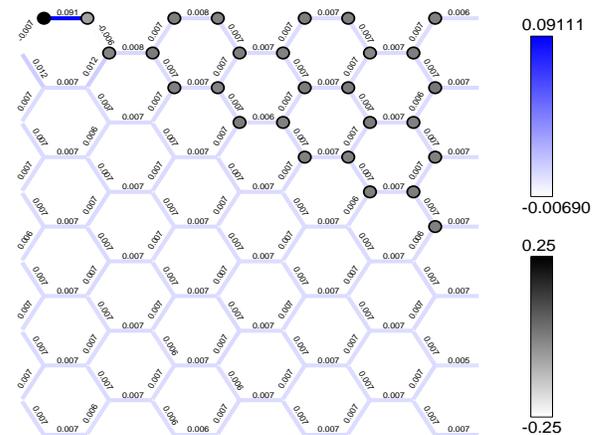}}
\caption{\label{f:bond} QMC results for correlation functions on an $N=288$ lattice. Each bond (and site) represents the strength of the correlation of a dimer (or spin) at the corresponding distance to the top left dimer (or spin) in a blue (dimers, top) or gray (spins, bottom) scale, respectively.}
\end{figure}

So far we only checked the disordered region for classical magnetic order. However, frustrated quantum models are known to exhibit in certain cases more exotic quantum ordered patterns such as dimer phases with long-range order.\cite{P:bartosch05, P:mosadeq11}
To check for this kind of order we calculated higher-order correlation functions of the spin variables, i.e., fourth-order correlations using improved estimators in the QMC simulations:\cite{P:sandvik92}
\begin{align}
\langle \vec S_i \vec S_j \vec S_k \vec S_l \rangle - \langle \vec S_i \vec S_j\rangle \langle \vec S_k \vec S_l \rangle\,,
\end{align}
where $i$ and $j$ index sites at one nearest-neighbor bond and $k$ and $l$ at another nearest-neighbor bond. 

\begin{figure}[!t]
\subfigure[\label{f:super_comp}ferromagnetic ($t=-0.5$) ]{\includegraphics[width=0.4\linewidth]{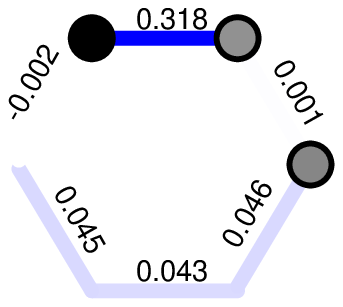}}\quad\quad
\subfigure[\label{f:dis_comp}disordered ($t=-0.1$)]{\includegraphics[width=0.4\linewidth]{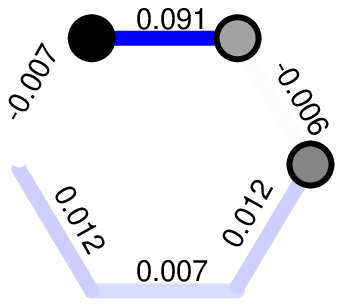}}
\caption{\label{f:corr_comp} Direct comparison of dimer correlations -- in an enlarged illustration of the top left hexagon of Figs.~\ref{f:bond}(b) and (c) -- shows a small relative enhancement of correlations on two bonds for the disordered phase ($V=0.45$ and original lattice $L=12$). The upper bond is again the reference bond for all dimer correlations.}
\end{figure}

Figure~\ref{f:bond} shows these correlations on a representative lattice where the strength of the correlation is given in a color code (blue scale, top) and the distance of the two bonds $i$-$j$ and $k$-$l$ is given by the distance of each bond to the top left reference bond. In addition, the $S^z$ correlation functions are shown in a gray scale (bottom scale) on the sites also with respect to the top left site of the lattice.
In Fig.~\ref{f:neelcorr} the values of these correlations are shown for a Néel-ordered state where the $S^z$ correlations oscillate for different sublattices and show a constant nearly maximal amplitude. The dimer correlations are small and show no sign of ordering. For parameters inside the in-plane ferromagnetic region [Fig.~\ref{f:supercorr}] neither in the $S^z$ spin nor in the dimer correlations can any signature be detected, i.e., spin correlations drop rapidly to zero and dimer correlations adopt a constant distance-independent value. The same applies for the disordered region [Fig.~\ref{f:discorr}] and only a minor detail distinguishes the two calculations: Apart from the different scales (cf.\,numbers at the upper scale), which are explained by the different strength of the quantum fluctuations ($t=-0.1,~-0.5$), we identify a small difference by comparing the relative values of the dimer correlations inside the top left hexagon shown enlarged in Fig.~\ref{f:corr_comp}.
In particular a slight enhancement of the dimer correlations on the bonds neighboring the opposite bond of the reference bond (top) compared to the correlation on the opposite bond itself is seen.
Thus an extremely short-ranged ordering of dimers is observed in the disordered phase, which is absent in the ferromagnetic state.

\begin{figure}[!t]
\includegraphics[width=0.48\textwidth]{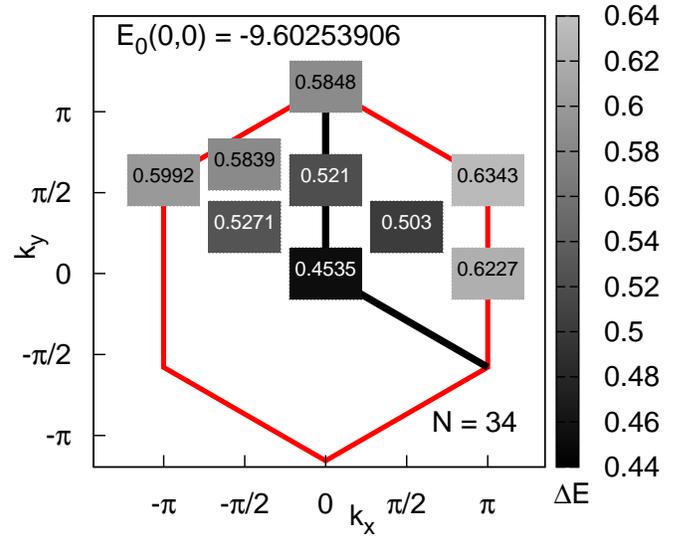}
\caption{\label{f:kspectrum} Spectrum of energy gaps $\Delta E_\vec{k} = E_\vec{k}-E_0$ for the $S^z_{\text{total}}=0$ subspace for $N=34$ at $V=0.45$ and $t=-0.1$. The smallest gap at $\vec k=0$ is stable in a finite-size analysis.}
\end{figure}

\begin{figure}[!h]
\includegraphics[width=0.48\textwidth]{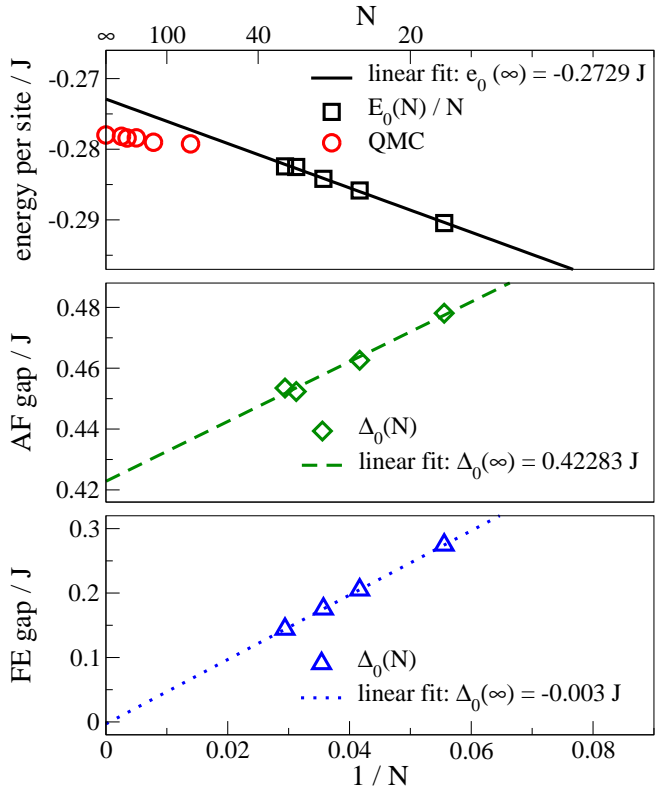}
\caption{\label{f:gaps} Scaling of the ground-state energy and two gaps with the system size $N$ from ED at $V=0.45$ and $t=-0.1$. The antiferromagnetic gap (AF, middle) inside the $S_{\text{total}}^z=0$ subspace stays finite; however, the ferromagnetic (FE, bottom) gap between the ground-state energy and the lowest eigenvalue of the $S_{\text{total}}^z=1$ subspace scales to zero.}
\end{figure}

Since we have found no finite signal for any magnetic or dimer order parameter we interpret the finite region in the phase diagram as disordered. To learn more about the character of this phase, an investigation of the low-energy spectrum is necessary. Therefore, we performed an ED of the lattice model for a set of parameters in the disordered region ($V=0.45$ and $t=-0.1$). 
In Fig.~\ref{f:kspectrum} we show the energy differences $\Delta E_\vec{k} = E_\vec{k}-E_0$ in the $S^z_{\text{total}}=0$ subspace. 
This measures the gap from the ground-state energy $E_0$ (which belongs to the $\vec k=0$ subspace and its scaling is shown in the top panel of Fig.~\ref{f:gaps}) to the lowest eigenvalues $E_\vec{k}$ in the different $\vec k\neq 0$ subspaces or to the first excited state in  the $\vec k=0$ subspace.
The smallest gap in this same subspace stays finite according to a finite-size analysis shown in Fig.~\ref{f:gaps} (middle, AF gap). 

We also calculated the lowest eigenvalues for higher $S^z_{\text{total}}$ subspaces and found that the gap between $E_0$ and the lowest eigenvalue in the $S^z_{\text{total}}=1$ subspace vanishes in a finite-size scaling analysis for system sizes $N = 18$ - $34$ (plotted in the bottom panel of Fig.~\ref{f:gaps}).
This indicates a ferromagnetic correlation, which can also be seen in the correlation functions of the $S^{xy}$ components in the ED.
However, from the scaling of the ferromagnetic gap it is obvious that the ED suffers from finite-size problems for the small lattices, which were investigated since we could exclude ferromagnetic order from the QMC simulations. 

In spite of finite-size effects, the clearly finite gap in the $S^z_{\text{total}}=0$ subspace found by ED in the disordered phase for the present model allows us to exclude a topologically ordered state. The reason is that, on a two-dimensional periodic lattice with an aspect ratio close to one (i.e., on a torus with similar circumferences in both directions), a topologically ordered state would be accompanied by a fourfold ground-state degeneracy.\cite{B:misguich2, *P:misguich02}

\section{Conclusion \label{s:dis}}
We presented the phase diagram for an anisotropic Heisenberg model on the honeycomb lattice with up to third-nearest-neighbor interactions. The quantum fluctuations in the present model are chosen to be ferromagnetic in order to avoid the sign problem in QMC simulations. Since apart from the interaction $J_2$ this sign can be absorbed by a sublattice rotation, one may hope to nevertheless gain insight into the isotropic model.

The frustrating next-nearest-neighbor interactions suppress the Néel state, which is favored by the antiferromagnetic $S^z$ interactions between nearest and third-nearest neighbors. Thus a collinear state arises for large interactions between the sites on the same sublattice. The schematic transition lines were calculated by means of LSWs and the direct transition between the competing antiferromagnetic states was found to survive for small fluctuations by means of SE. In addition, we also performed QMC simulations to determine the phase diagram and in particular to gain insight into a region of the phase diagram without any conventional order. Since an earlier ED study of the isotropic model reported the appearance of valence bond crystals\cite{P:albuquerque11} for certain parameters we calculated higher-order correlation functions and excluded any long-range order of dimer configurations. This absence of any finite order parameter is in agreement with earlier calculations for the isotropic model,\cite{P:cabra10, P:albuquerque11, P:oitmaa11} although the stability region of this disordered phase differs at a quantitative level. A further investigation of the low-energy spectrum using ED was hampered by finite-size effects but yielded a finite antiferromagnetic gap, which indicates absence the of topological order. For isotropic models the disordered state with a finite gap would be referred to as a gapped spin-liquid phase. In principle, similar finite-size effects as in ED are also present in the QMC simulations, but here they can be overcome by calculating larger systems. This underlines again the utility of introducing ferromagnetic quantum fluctuations.

The phase diagram is very similar to the one found for the square lattice with nearest- and next-nearest-neighbor interactions, which was investigated for anisotropic exchange terms in Ref.~\onlinecite{P:kalz11}. However, the stability region of the antiferromagnetic phases is reduced on the honeycomb lattice, which is explained by the lower coordination number of the lattice, thus the influence of quantum fluctuations is enhanced. In particular, the critical value of the ratios $t_r/V_r$ obtained by QMC simulations at which the antiferromagnetic phase boundaries split and the disordered phase emerges is smaller in the present case. Furthermore, the direct transition line between the two antiferromagnetic states calculated via SE shows a steeper slope for the square lattice which also hints at a larger stability of this direct transition in that case.\cite{P:kalz11} 

\begin{acknowledgments}
We would like to thank the DFG for financial support via the Collaborative Research Center SFB 602 (TP A18) and a Heisenberg fellowship (Grant No.\,HO 2325/4-2, A.\ H.). The collaboration with Argentina was also funded by the DAAD via a short-term scholarship (Grant No. D/10/46833, A.\ K.). M.\ A., D.\ C., and G.\ R. were partially supported by CONICET (PIP 1691) and ANPCyT (PICT 1426).
Furthermore, we mention that most of the QMC simulations were performed on the parallel clusters of the North-German Supercomputing Alliance (HLRN) and thank them for technical support. Finally, we would like to thank Jörg Schulenburg for providing us with his diagonalization code.
\end{acknowledgments}
\pagebreak

\bibliography{Literatur_hex}

\end{document}